\documentclass[preprint,showpacs,eqsecnum,epsf,aps,floats,tighten]{revtex4}
\usepackage{graphicx}  
\usepackage{dcolumn}   
\usepackage{bm}        
\usepackage{amssymb}   

\def\CLB{CL_{\it B}}
\def\CLS{CL_{\it S}}
\def\CLSB{CL_{\it S+B}}
\def\Hpm{ {\it H}^{\pm\pm}}
\def\HpmLR{ {\it H}^{\pm\pm}_{\it L(R)}}
\def\HpmL{ {\it H}^{\pm\pm}_{\it L}}
\def\HpmR{ {\it H}^{\pm\pm}_{\it R}}

\def\Hpp{ {\it H}^{++}}
\def\Hmm{ {\it H}^{--}}
\def\ppbar{{\it p}\overline{\it p}}
\def\ttbar{{\it t}\overline{\it t}}
\def\bbbar{{\it b}\overline{\it b}}
\def\qqbar{{\it q}\overline{\it q}}

\begin{document}

\title{
Search for Doubly-charged Higgs Boson Pair Production in the
Decay to~$\bm{\mu^{+}\mu^{+}\mu^{-}\mu^{-}}$  
in $\bm\ppbar$ Collisions at $\bm{\sqrt{s}=1.96}$~TeV}

%
\author{                                                                      
V.M.~Abazov,$^{32}$                                                           
B.~Abbott,$^{69}$                                                             
M.~Abolins,$^{60}$                                                            
B.S.~Acharya,$^{26}$                                                          
D.L.~Adams,$^{67}$                                                            
M.~Adams,$^{47}$                                                              
T.~Adams,$^{45}$                                                              
M.~Agelou,$^{16}$                                                             
J.-L.~Agram,$^{17}$                                                           
S.N.~Ahmed,$^{31}$
% ...
\\                                                                            
\vskip 0.30cm                                                                 
\centerline{(D\O\ Collaboration)}                                             
\vskip 0.30cm                                                                 
}                                                                             
\address{                                                                     
\centerline{$^{1}$Universidad de Buenos Aires, Buenos Aires, Argentina}       
\centerline{$^{2}$LAFEX, Centro Brasileiro de Pesquisas F{\'\i}sicas,         
                  Rio de Janeiro, Brazil}                                     
\centerline{$^{3}$Universidade do Estado do Rio de Janeiro,                   
                  Rio de Janeiro, Brazil}                                     
\centerline{$^{4}$Instituto de F\'{\i}sica Te\'orica, Universidade            
                  Estadual Paulista, S\~ao Paulo, Brazil}                     
\centerline{$^{5}$University of Alberta and Simon Fraser University,          
                  Canada}                                                     
\centerline{$^{6}$Institute of High Energy Physics, Beijing,                  
                  People's Republic of China}                                 
\centerline{$^{7}$Universidad de los Andes, Bogot\'{a}, Colombia}             
\centerline{$^{8}$Charles University, Center for Particle Physics,            
                  Prague, Czech Republic}                                     
\centerline{$^{9}$Czech Technical University, Prague, Czech Republic}         
\centerline{$^{10}$Institute of Physics, Academy of Sciences, Center          
                  for Particle Physics, Prague, Czech Republic}               
\centerline{$^{11}$Universidad San Francisco de Quito, Quito, Ecuador}        
\centerline{$^{12}$Laboratoire de Physique Subatomique et de Cosmologie,      
                  IN2P3-CNRS, Universite de Grenoble 1, Grenoble, France}     
\centerline{$^{13}$CPPM, IN2P3-CNRS, Universit\'e de la M\'editerran\'ee,     
                  Marseille, France}                                          
\centerline{$^{14}$Laboratoire de l'Acc\'el\'erateur Lin\'eaire,              
                  IN2P3-CNRS, Orsay, France}                                  
\centerline{$^{15}$LPNHE, Universit\'es Paris VI and VII, IN2P3-CNRS,         
                  Paris, France}                                              
\centerline{$^{16}$DAPNIA/Service de Physique des Particules, CEA, Saclay,    
                  France}                                                     
\centerline{$^{17}$IReS, IN2P3-CNRS, Univ. Louis Pasteur Strasbourg,          
                   and Univ. de Haute Alsace, France}                         
\centerline{$^{18}$Institut de Physique Nucl\'eaire de Lyon, IN2P3-CNRS,      
                   Universit\'e Claude Bernard, Villeurbanne, France}         
\centerline{$^{19}$RWTH Aachen, III. Physikalisches Institut A,               
                   Aachen, Germany}                                           
\centerline{$^{20}$Universit{\"a}t Bonn, Physikalisches Institut,             
                  Bonn, Germany}                                              
\centerline{$^{21}$Universit{\"a}t Freiburg, Physikalisches Institut,         
                  Freiburg, Germany}                                          
\centerline{$^{22}$Universit{\"a}t Mainz, Institut f{\"u}r Physik,            
                  Mainz, Germany}                                             
\centerline{$^{23}$Ludwig-Maximilians-Universit{\"a}t M{\"u}nchen,            
                   M{\"u}nchen, Germany}                                      
\centerline{$^{24}$Fachbereich Physik, University of Wuppertal,               
                   Wuppertal, Germany}                                        
\centerline{$^{25}$Panjab University, Chandigarh, India}                      
\centerline{$^{26}$Tata Institute of Fundamental Research, Mumbai, India}     
\centerline{$^{27}$University College Dublin, Dublin, Ireland}                
\centerline{$^{28}$Korea Detector Laboratory, Korea University,               
                   Seoul, Korea}                                              
\centerline{$^{29}$CINVESTAV, Mexico City, Mexico}                            
\centerline{$^{30}$FOM-Institute NIKHEF and University of                     
                  Amsterdam/NIKHEF, Amsterdam, The Netherlands}               
\centerline{$^{31}$University of Nijmegen/NIKHEF, Nijmegen, The               
                  Netherlands}                                                
\centerline{$^{32}$Joint Institute for Nuclear Research, Dubna, Russia}       
\centerline{$^{33}$Institute for Theoretical and Experimental Physics,        
                  Moscow, Russia}                                             
\centerline{$^{34}$Moscow State University, Moscow, Russia}                   
\centerline{$^{35}$Institute for High Energy Physics, Protvino, Russia}       
\centerline{$^{36}$Petersburg Nuclear Physics Institute,                      
                   St. Petersburg, Russia}                                    
\centerline{$^{37}$Lund University, Royal Institute of Technology,            
                   Stockholm University, and Uppsala University, Sweden}      
\centerline{$^{38}$Lancaster University, Lancaster, United Kingdom}           
\centerline{$^{39}$Imperial College, London, United Kingdom}                  
\centerline{$^{40}$University of Manchester, Manchester, United Kingdom}      
\centerline{$^{41}$University of Arizona, Tucson, Arizona 85721}              
\centerline{$^{42}$Lawrence Berkeley National Laboratory and University of    
                  California, Berkeley, California 94720}                     
\centerline{$^{43}$California State University, Fresno, California 93740}     
\centerline{$^{44}$University of California, Riverside, California 92521}     
\centerline{$^{45}$Florida State University, Tallahassee, Florida 32306}      
\centerline{$^{46}$Fermi National Accelerator Laboratory, Batavia,            
                   Illinois 60510}                                            
\centerline{$^{47}$University of Illinois at Chicago, Chicago,                
                   Illinois 60607}                                            
\centerline{$^{48}$Northern Illinois University, DeKalb, Illinois 60115}      
\centerline{$^{49}$Northwestern University, Evanston, Illinois 60208}         
\centerline{$^{50}$Indiana University, Bloomington, Indiana 47405}            
\centerline{$^{51}$University of Notre Dame, Notre Dame, Indiana 46556}       
\centerline{$^{52}$Iowa State University, Ames, Iowa 50011}                   
\centerline{$^{53}$University of Kansas, Lawrence, Kansas 66045}              
\centerline{$^{54}$Kansas State University, Manhattan, Kansas 66506}          
\centerline{$^{55}$Louisiana Tech University, Ruston, Louisiana 71272}        
\centerline{$^{56}$University of Maryland, College Park, Maryland 20742}      
\centerline{$^{57}$Boston University, Boston, Massachusetts 02215}            
\centerline{$^{58}$Northeastern University, Boston, Massachusetts 02115}      
\centerline{$^{59}$University of Michigan, Ann Arbor, Michigan 48109}         
\centerline{$^{60}$Michigan State University, East Lansing, Michigan 48824}   
\centerline{$^{61}$University of Mississippi, University, Mississippi 38677}  
\centerline{$^{62}$University of Nebraska, Lincoln, Nebraska 68588}           
\centerline{$^{63}$Princeton University, Princeton, New Jersey 08544}         
\centerline{$^{64}$Columbia University, New York, New York 10027}             
\centerline{$^{65}$University of Rochester, Rochester, New York 14627}        
\centerline{$^{66}$State University of New York, Stony Brook,                 
                   New York 11794}                                            
\centerline{$^{67}$Brookhaven National Laboratory, Upton, New York 11973}     
\centerline{$^{68}$Langston University, Langston, Oklahoma 73050}             
\centerline{$^{69}$University of Oklahoma, Norman, Oklahoma 73019}            
\centerline{$^{70}$Brown University, Providence, Rhode Island 02912}          
\centerline{$^{71}$University of Texas, Arlington, Texas 76019}               
\centerline{$^{72}$Rice University, Houston, Texas 77005}                     
\centerline{$^{73}$University of Virginia, Charlottesville, Virginia 22901}   
\centerline{$^{74}$University of Washington, Seattle, Washington 98195}       
}                                                                             
\date{\today}

\begin{abstract}
A search for pair production of 
doubly-charged Higgs bosons in the process 
$\ppbar\to\Hpp\Hmm\to \mu^+\mu^+\mu^-\mu^-$ is performed
with the D\O\ Run II detector at the Fermilab Tevatron.
The analysis is based on a sample of inclusive di-muon 
data collected at an energy of $\sqrt{s}=1.96$~TeV,
corresponding to an integrated luminosity of $113~$pb$^{-1}$.
In the absence of a signal, $95\%$~confidence level
mass limits of $M(\HpmL)>118.4$~GeV$/c^2$ and $M(\HpmR)>98.2$~GeV$/c^2$ 
are set for left-handed and right-handed
doubly-charged Higgs bosons, respectively, assuming $100\%$ branching
into muon pairs.
\end{abstract}

\pacs{14.80.Cp
\hspace{2cm}  FERMILAB-Pub-04/045-E}
\maketitle
Doubly-charged Higgs bosons appear in theories
beyond the Standard Model, 
in particular, in left-right symmetric models~\cite{bib-lr}, 
in Higgs triplet models~\cite{bib-theo} and in 
Little-Higgs models~\cite{bib-little}. 
The models predict dominant decay modes to like-charge lepton pairs,
$\Hpm\to\ell^{\pm}\ell^{\pm}$.
Pairs of doubly-charged Higgs bosons 
can be produced through the Drell-Yan process
$\qqbar\to\gamma^{*}/{\it Z}\to\Hpp\Hmm$.
Next-to-leading order (NLO) corrections to this cross section 
have recently been calculated~\cite{bib-spira}. 
The pair production cross sections 
for left-handed states in the mass
range studied in this Letter are about a factor two 
larger than for the right-handed
states due to different coupling to the {\it Z} boson.
Left-handed and right-handed states are distinguished through their
decays into left-handed or right-handed leptons. 
The cross section also depends on the hypercharge $Y$
of the $\Hpm$ boson.

The $\Hpm$ decay width into leptons
is given by $\Gamma^{\ell\ell}=(8\pi)^{-1}|h_{\ell\ell}|^2 
M_{\Hpm}$, where $h_{\ell\ell}$ is the Yukawa coupling to 
leptons~\cite{bib-theo}. 
A limit on $h_{\mu\mu}^2/M_{\Hpm}^2$, where $h_{\mu\mu}$ is
the Yukawa coupling to muons, can be
derived from the expected contribution to the anomalous
magnetic moment of the muon, $(g-2)_{\mu}$~\cite{bib-huitu}.
This yields upper limits on the
Yukawa coupling $h_{\mu\mu}$ of the order 0.1 
for $M_{\Hpm }= 100$~GeV$/c^2$. 
Requiring that $\Hpm$ bosons decay 
within about $1$~cm of their production restricts
the sensitivity to $h_{\mu\mu}$ to
approximately greater than $10^{-7}$.

Experiments at the CERN LEP collider
have searched for pair production of doubly-charged 
Higgs bosons in ${\it e}^+{\it e}^-$ interactions. 
Mass limits for decays into muons of $M(\HpmL) > 100.5$~GeV$/c^2$ 
and $M(\HpmR) > 100.1$~GeV$/c^2$ were obtained by the OPAL 
collaboration~\cite{bib-OPAL}, 
and a limit of $M(\HpmLR) > 99.4$~GeV$/c^2$ 
by the L3 collaboration~\cite{bib-L3}.
Similar limits were set for decays into electrons~\cite{bib-OPAL,bib-L3}
and $\tau$-leptons~\cite{bib-OPAL,bib-L3,bib-DELPHI}.
Our measurement represents the first $\Hpm$ search at hadron colliders, and
it extends significantly the range of sensitivity
for left-handed doubly-charged Higgs bosons decaying into muons.
All limits in this Letter are given at $95\%$ 
confidence level (CL).

The D\O\ Run II detector comprises 
a central tracking system, a
liquid-argon/uranium calorimeter, and an iron toroid muon
spectrometer~\cite{run2det}.
The central tracking system 
consists of a silicon microstrip tracker
(SMT) and a central fiber tracker (CFT), both located within a 2~T 
superconducting solenoidal magnet. 
The SMT strips have a typical pitch 
of $50$--$80$~$\mu$m, and a design optimized for tracking and vertexing 
capability in the pseudorapidity range $|\eta|<3$.
The system has a six-barrel longitudinal structure, 
each with a set of four layers arranged axially around the beam pipe, 
and interspersed with sixteen 
radial disks. The CFT has eight thin coaxial barrels, 
each supporting two doublets of overlapping scintillating fibers of 0.835~mm 
diameter, one doublet being parallel to the collision axis, and the other 
alternating by $\pm 3^{\circ}$ relative to the axis. 
The calorimeters consist of 
a central section (CC) covering $|\eta|$
up to $\approx 1$, and 
two end calorimeters (EC) extending 
coverage to $|\eta|<4.2$, 
all housed in separate cryostats~\cite{run1det}. 
Scintillators between the 
CC and EC cryostats provide sampling of showers at $1.1<|\eta|<1.4$.
A muon system resides beyond the calorimetry, and 
consists of a layer of tracking 
detectors and scintillation counters before 1.8~T iron toroids, 
followed by two more similar layers after the toroids. 
Tracking at $|\eta|<1$ 
relies on 10~cm wide drift tubes~\cite{run1det}, while 1~cm mini-drift 
tubes are used at $1<|\eta|<2$. 

This analysis~\cite{bib-marian} is based on inclusive di-muon data recorded 
between August 2002 and June 2003. The events are triggered
by requiring two muon candidates in the muon scintillation
counters and at least one reconstructed muon using 
the muon wire chambers. 
The integrated luminosity is measured using two scintillator hodoscopes located
on either side of the interaction region.
For the di-muon triggers the integrated luminosity is $113\pm 7$~pb$^{-1}$. 
 
Event selection proceeds in four steps.
The first step (selection S1) requires at least two muons.
Each muon used in the analysis must have a 
transverse momentum $p_{\it T}>15$~GeV$/c$,
where $p_{\it T}$ is measured with respect to the beam axis.
The muon tracks are reconstructed using 
wire and scintillator hits in the 
different layers of the muon system.
These must be combined
successfully with a central track reconstructed in the SMT and CFT
detectors to measure the muon momentum. 
A requirement on the timing of
hits in different scintillator layers is used to
minimize background from cosmic rays.

The second set of selections (S2) is based on 
isolation criteria based on calorimeter and tracking information,
and is designed primarily to reject background 
from muons originating from semi-leptonic {\it B} hadron decays. 
The direction of each muon track is projected through
the calorimeter. For at least two muons,
the sum of the transverse energies of the calorimeter cells in an annular
ring $0.1<R<0.4$ around each muon direction is required to be
$\sum_{\text{cells},i} E_{\it T}^i< 2.5$~GeV, 
where $R=\sqrt{(\Delta\phi)^2+(\Delta\eta)^2}$ and $\phi$ is
the azimuthal angle.
In addition, the sum of the transverse momenta of all tracks
other than that of the muon in a cone of $R=0.5$ around the muon track
is required to satisfy
$\sum_{{\text{tracks}},i} p_{\it T}^i< 2.5$ GeV$/c$.

Selection (S3) applies to events with just two muons and
requires a difference in azimuthal 
angle $\Delta\phi<0.8\pi$. 
It is applied to reject ${\it Z}\to\mu^+\mu^-$ events and 
to reduce background from semi-leptonic {\it B} hadron decays.
This selection also removes the remaining background from cosmic muons.

The final selection (S4) requires at least one pair of muons in the event 
to be of like-sign charge. These pairs are considered
candidates for $\Hpm\to\mu^{\pm}\mu^{\pm}$ decays. 

The geometric and kinematic acceptance is taken from a
{\sc geant}-based~\cite{bib-geant} simulation of the detector.
All other efficiencies are determined from ${\it Z}\to\mu^+\mu^-$ data.
The single muon detection and reconstruction efficiencies, and the
efficiency of the isolation requirement
are measured by using one muon to tag the event
and the second muon to measure the efficiencies.
The trigger efficiency is measured
by analyzing events with calorimeter-based triggers which
are independent of the muon system.
To obtain the signal and background rates, 
corrections are applied to the simulation so that it
reproduces the measured efficiencies.
The total signal efficiency for our event selection is  
$(47.5\pm2.5)\%$ and does not depend on the mass of the 
doubly-charged Higgs in the mass range studied.
\begin{table}[htbp]
\caption{\label{tab-all}
Expected number of events 
for a signal with $M(\HpmL)=100$~GeV$/c^2$,
background events from Monte Carlo
for the available integrated data luminosity, and number of observed events
remaining after each selection cut.
The simulation of {\it Z} decays includes the Drell-Yan contribution.
}
\begin{ruledtabular}
\begin{tabular}{lcccc}
Selection   & 2 muons & Isolation  & $\Delta \phi<0.8\pi$  & Like-sign\\
            & $p_T>15$~GeV$/c$   &  &   &  \\
            & S1  & S2 &  S3 &  S4\\
\hline  
Signal     & $9.4$      & $8.5$  &  $7.5$  & $6.5$      \\
\hline
${\it Z}
\to\mu^+\mu^- $       & $4816$     & $4055$          &  $359$      & $0.3\pm0.1$\\ 
$ \bbbar $            & $391$  & $18$  &  $3.0$  & $0.8\pm0.4$\\ 
$ {\it Z}\to \tau^+\tau^- $    & $40$  & $34$   &  $2.4$    & $<0.1$     \\  
$ \ttbar $            & $3.0$  & $2.1$ & $1.5$  & $0.11\pm0.01$     \\
{\it ZZ}              & $0.1$  & $0.1$ & $0.1$  & $0.05\pm0.01$     \\
{\it WZ}              & $0.6$  & $0.5$ & $0.4$  & $0.23\pm0.01$     \\
{\it WW}              & $3.5$  & $3.1$ & $1.9$  & $<0.01$ \\
\hline
Total & & & & \\
background            & $5254 \pm 47$  & $4113 \pm 43$  &  $368 \pm 14$  &  $1.5\pm0.4$\\ 

\hline
Data                  & $5168$ & $4133$ & $378$ & $3$ \\ 
\end{tabular}
\end{ruledtabular}
\end{table}
\begin{table}[htbp]
\caption{\label{tab-like}
Expected number
of background events from Monte Carlo for the available data
luminosity, and number of observed events
remaining after each selection cut, with 
selection S4, requiring at least one like-charge muon pair, 
applied together with S1.
Only statistical uncertainties are given.
The contribution from {\it WW} events is negligible.}
\begin{ruledtabular}
\begin{tabular}{lccc}
Selection   & 2 muons & Isolation  & $\Delta \phi<0.8\pi$ \\
(Like-sign) & $p_T>15$~GeV$/c$         &   &  \\
            & S4 \& S1  & S2 &  S3 \\
\hline  
$Z\rightarrow\mu^+\mu^- $       & $0.9 \pm 0.3$   & $0.6\pm 0.2$ & $0.3\pm 0.1$ \\
$ \bbbar $                      & $95.1\pm 3.3$      & $4.4\pm 1.9$ & $0.8\pm 0.4$ \\
$ Z
\to \tau^+\tau^- $   &  $<0.1$&  $<0.1$&  $<0.1$ \\
$ \ttbar $                      & $0.24 \pm 0.01$ & $0.11\pm 0.01$ & $0.11\pm 0.01$ \\
{\it ZZ}                             & $0.06 \pm 0.01$ & $0.05\pm 0.01$ & $0.05\pm 0.01$ \\
{\it WZ}                            & $0.29 \pm 0.01$ & $0.27\pm 0.01$ & $0.23\pm 0.01$ \\
\hline
Total & & &  \\
background                        & $96.6\pm 3.3$   & $5.4\pm 1.9$ & $1.5\pm 0.4$ \\
\hline
Data                            & $101$ & $5$ & $3$ \\ 
\end{tabular}
\end{ruledtabular}
\end{table}

\begin{figure}[htbp]
\includegraphics[width=1.0\columnwidth]{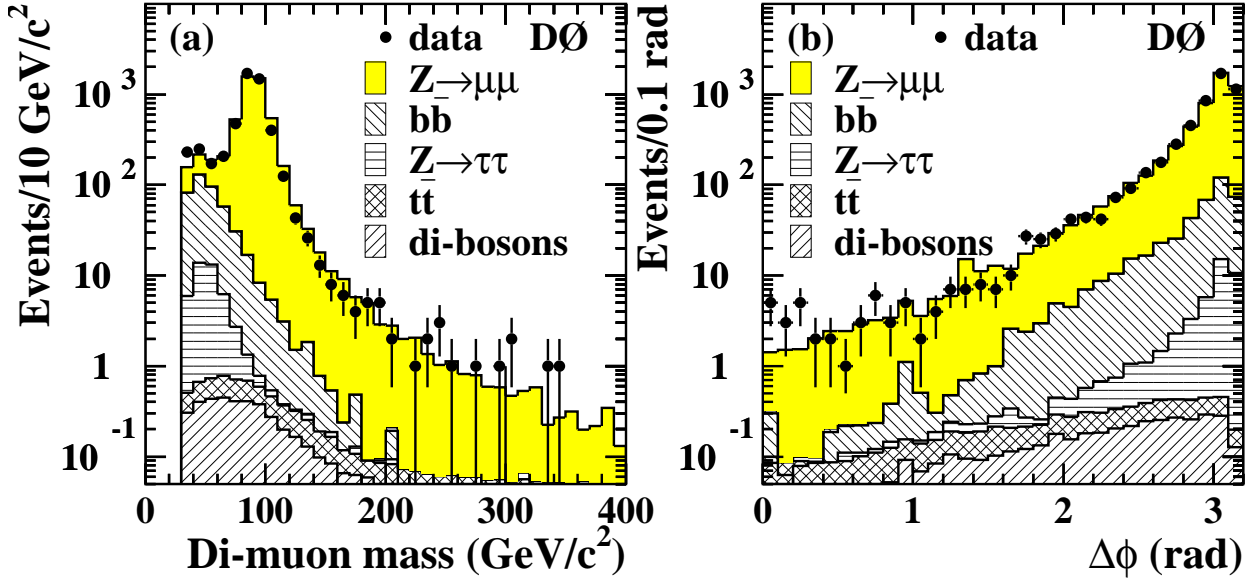}
\caption{\label{fig-cut1}
Measured distribution of (a) di-muon mass and 
(b) $\Delta\phi$ compared 
to the sum of MC background processes after selection S1.
Di-muon mass and $\Delta\phi$ are calculated for the two 
highest $p_{\it T}$ muons, independent of their charge.
}
\end{figure}

Distributions of the di-muon mass and of $\Delta\phi$ after 
selection S1 are shown in Fig.~\ref{fig-cut1}.
The data are compared to the sum
of Monte Carlo (MC) contributions from different
background processes. 
In events with more than two muons, the 
di-muon mass and $\Delta\phi$ are calculated only for the 
two muons of highest $p_{\it T}$, independent of their charge.
Both signal and background events are generated with 
{\sc pythia~6.2}~\cite{bib-pythia}.
The NNLO cross-section is used to normalize the
${\it Z}\to \mu^+\mu^-$ sample~\cite{bib-nlo}.
{\sc pythia} does not provide a good description of the jet
multiplicity in {\it Z}+jets events.
Since the $\Delta\phi$ distributions
are expected to be sensitive to the number of jets,
the simulated ${\it Z}\to \mu^+\mu^-$ events are
re-weighted to reproduce the distribution of
jet multiplicities observed in data.
There is agreement between data and MC simulation, 
for both the normalization and shapes of the di-muon mass and
$\Delta\phi$ distributions (Fig.~\ref{fig-cut1}). 
The number of events remaining after each selection
and the efficiencies for a signal
of a mass of $M(\HpmL)=100$~GeV$/c^2$ are given in Table~\ref{tab-all}.

The background contribution from
$\ttbar$ and di-boson ({\it WZ, ZZ} and {\it WW}) production
is also estimated by MC simulation.
The NLO cross section is used for $\ttbar$ events~\cite{bib-nlott}.
Higher-order QCD corrections to di-boson production are approximated
by multiplying the LO cross section given in
{\sc pythia} by a $K$-factor of $1.34$.

When the requirement of having at least one pair of like-charge muons is 
applied simultaneously with S1, most of the background
from {\it Z} boson decays is removed, and only 101 like-sign events remain 
(Table~\ref{tab-like}). Since no isolation requirement is imposed
at this stage, the largest remaining background is from $\bbbar$
production.
{\sc pythia} is used to estimate this background 
by generating inclusive jet events with
a minimum transverse momentum 
of $30$~GeV$/c$ for the hard interaction.
The inclusive {\it b} quark production cross section 
$\sigma^{\it b}(p_{\it T}^{\it b}>30\mbox{~GeV$/c$})$ was
measured by D\O\ to be $54\pm 20$~nb in the rapidity interval 
$|y^{\it b}|<1$ at $\sqrt{s}=1.8$~TeV~\cite{bib-run1}. 
This cross section is extrapolated via {\sc pythia} to the full $y^{\it b}$
range and to $\sqrt{s}=1.96$~TeV, and is then used to normalize the
$\bbbar$ MC sample.

\begin{figure}[htbp]
\includegraphics[width=1.0\columnwidth]{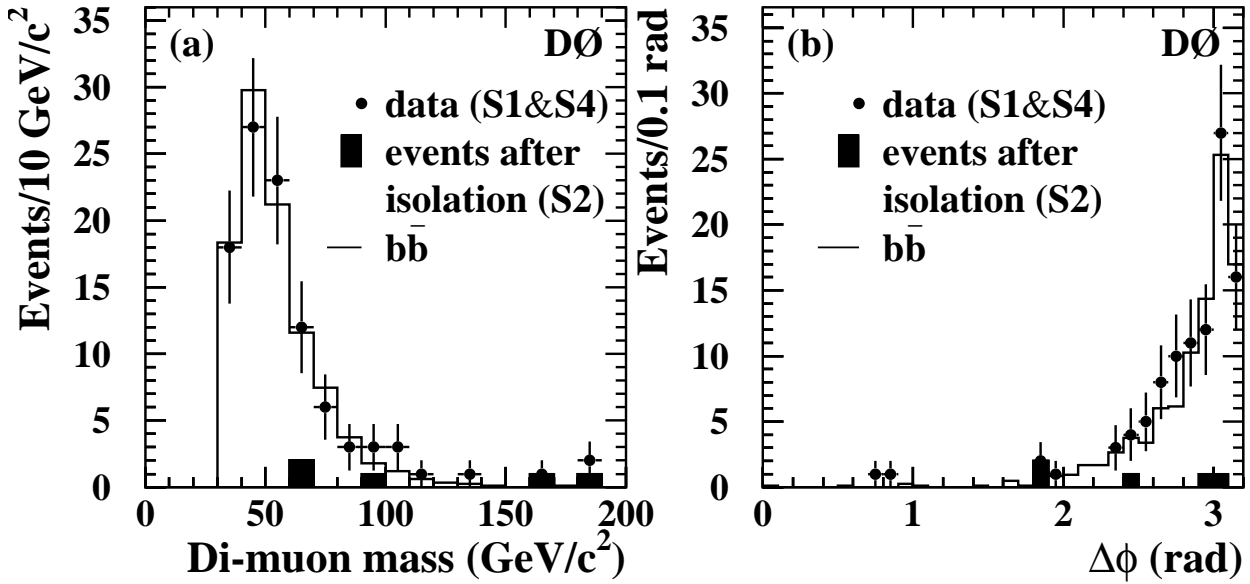}
\caption{\label{fig-like}
(a) Di-muon mass and (b) $\Delta\phi$ for the two like-charge muons 
with the highest $p_{\it T}$. The 101 events remaining in the data 
after the selections S1 and S4 (points with error bars) are compared to 
the {\sc pythia} $\bbbar$ simulation (open histogram).
The five data events remaining after the isolation selection are shown 
separately (full histogram).
}
\end{figure}

Distributions in di-muon mass and $\Delta\phi$ 
for the like-sign events are compared to the {\sc pythia} $\bbbar$ simulation
in Fig.~\ref{fig-like}. Since data and Monte Carlo are in good agreement,
the efficiency of the $\Delta\phi$ requirement is taken from the simulation.
The data are used to determine the isolation efficiency.
Out of 101 like-sign events, five remain after applying the isolation 
requirement (S2).
Assuming that all like-sign events originate from $\bbbar$ processes,
the isolation efficiency  for $\bbbar$ events is found to be $(5\pm2)\%$, and
the background from $\bbbar$ production
in the final sample is expected to be $0.8\pm0.3$ events.
Adding a systematic uncertainty of $37\%$ 
on the measured $\bbbar$ cross section~\cite{bib-run1}
yields a total uncertainty on the $\bbbar$ background of $50\%$.

Another potential background is from
${\it Z}\to\mu^+\mu^-$ decays which are
not rejected by the $\Delta\phi<0.8\pi$ requirement,
and in which one of the
muon charges is misidentified. 
For very high $p_{\it T}$ tracks,
the uncertainty on the measured curvature 
can cause such a flip
of the track curvature. Most of these tracks have $|\eta|>1.62$, because
there are fewer CFT layers in this region. 
The ${\it Z}\to\mu^+\mu^-$ simulation predicts
$0.3\pm0.1$ events after the final selection.
We have also estimated the probability
for charge misidentification using data. 
The upper limit is given by the
ratio of like-sign (5) to opposite sign (4133) events after the selection S2 
(Tables~\ref{tab-all} and \ref{tab-like}), and equals $0.12\%$.
Since 378 events remain before the like-sign requirement, then
assuming that the charge-misidentification probability is independent of the
$\Delta\phi$ requirement, less than $0.5\pm0.2$ 
background events are expected due
to charge misidentification. This is in good agreement with the simulation.
\begin{figure}[htbp]
\includegraphics[width=1.0\columnwidth]{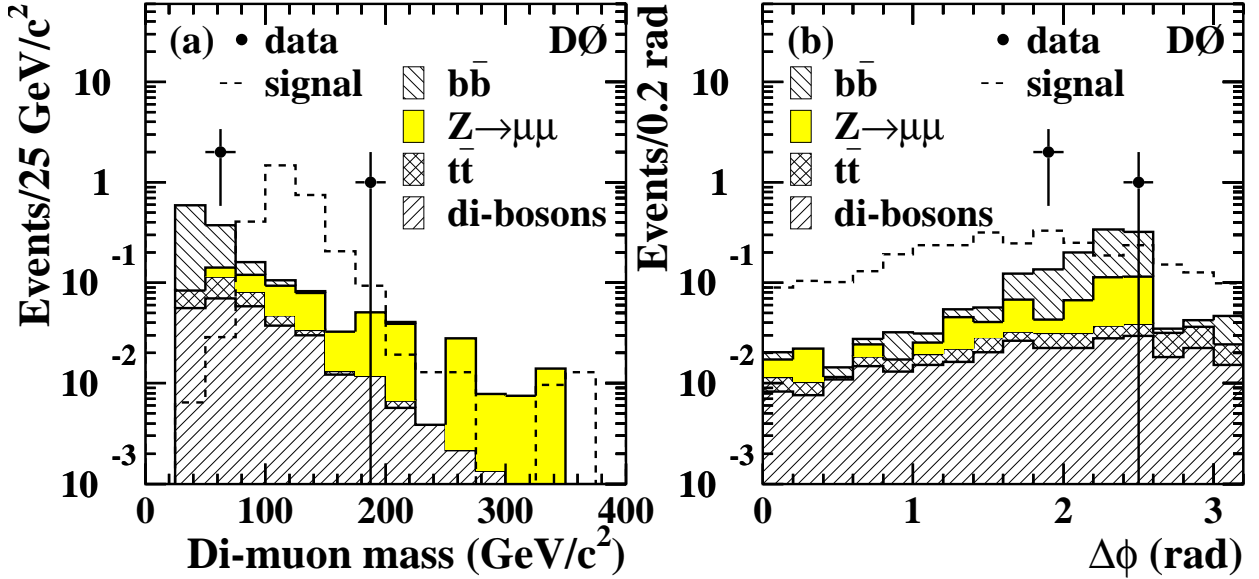}
\caption{\label{fig-cut2}
(a) Di-muon mass and (b) $\Delta\phi$ for the two like-charge muons
of  highest $p_{\it T}$. 
The data are compared to the sum of MC background processes after 
all selections.
The open histogram shows the distributions
for a left-handed, doubly-charged Higgs boson with a mass of $120$~GeV$/c^2$.
Since the $\Delta\phi$ requirement is only applied
to events with two muons, events with more than two muons 
can contribute for $\Delta\phi>0.8\pi$. 
}
\end{figure}

The production of {\it W} bosons decaying into 
$\mu\nu$, in association with jets,
is another source of background, but mainly at low di-muon mass.
By extrapolating to $p_{\it T}>15$~GeV$/c$  
the steeply falling $p_{\it T}$ spectrum of muons that fail the isolation 
requirements in di-muon events from a sample of ${\it W}\to\mu\nu$+jets data,
we estimate this contribution to be less than $0.1$ events.
The expected background rate,
as determined from the data, is in agreement with the MC simulation.

Three candidates remain in the data after the final selection.
The di-muon mass and $\Delta\phi$ distributions for these
events are compared to the sum of MC backgrounds
in Fig.~\ref{fig-cut2}.
Two events have two negatively-charged
muons and one positively-charged muon.
Of the two, one has $\Delta\phi=2.48$, and it has the highest 
like-sign di-muon mass ($183$~GeV$/c^2$) of the three candidates.
The like-sign di-muon mass of the second event is $63$~GeV$/c^2$ and
the invariant mass of the two highest $p_{\it T}$ muons 
of opposite charge in this event is 91~GeV$/c^2$.
The third event has two positively-charged muons with a mass of $62$~GeV$/c^2$.
The higher $p_{\it T}$ track may be interpreted as a 
case of charge misidentification since it traverses the CFT in a region with
less than 16 layers.

\begin{figure}[htbp]
\includegraphics[width=0.98\columnwidth]{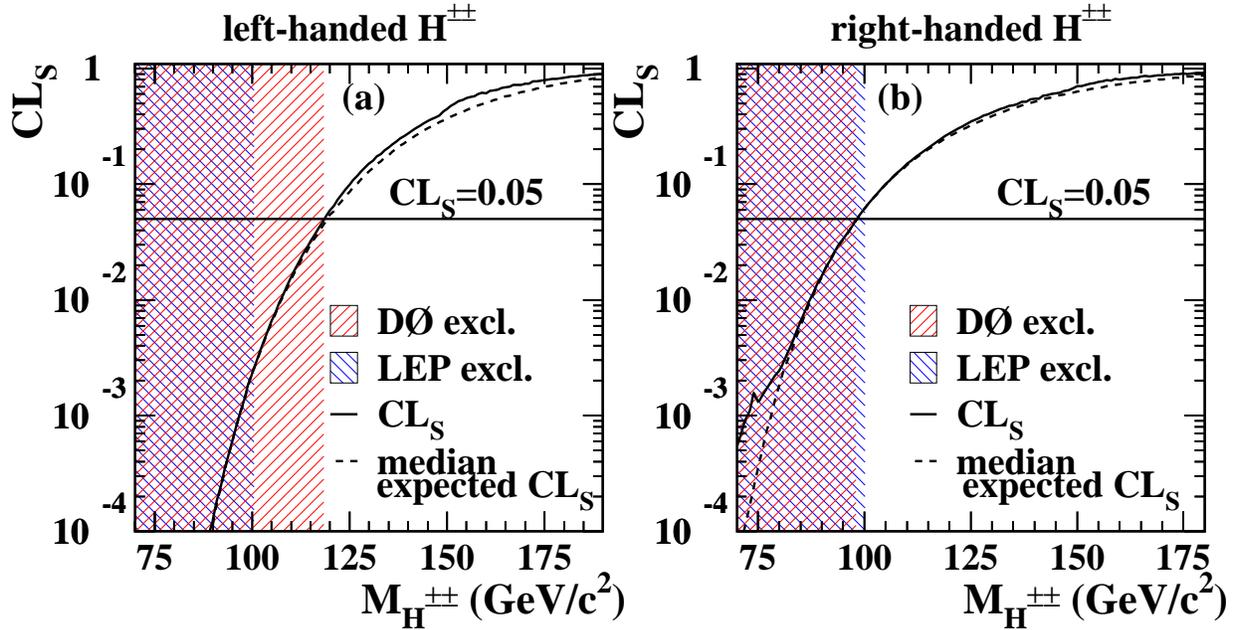}
\caption{\label{fig-cl}
Confidence level $\CLS$ as a function
of the mass $M_{\Hpm}$ for (a) left-handed and (b)
right-handed doubly-charged Higgs bosons. 
Masses with $\CLS<0.05$ are excluded by this analysis.
The mass regions excluded by LEP are also shown.
The impact of systematic uncertainties is included in the limits. 
The dashed curve shows median expected $\CLS$ for no signal.
}
\end{figure}
Since the remaining candidate events are consistent with
a background observation, $\Hpm$ mass limits
are derived using the program 
{\sc mclimit}~\cite{bib-tom}. It provides the confidence level for
the background to represent the data, $\CLB$, and the confidence level
for the sum of signal and background hypothesis, $\CLSB$,~\cite{bib-limit} 
taking into account the expected mass distribution for signal and 
background, and the mass resolution.
The mass resolution varies between
$\approx 7.6$~GeV$/c^2$ for $M_{\Hpm}=80$~GeV$/c^2$ and $\approx 30$~GeV$/c^2$ 
for $M_{\Hpm}=200$~GeV$/c^2$.
The expected rate for signal as a function of the Higgs mass 
is determined by the NLO cross section~\cite{bib-spira}, 
the signal efficiencies, and the measured luminosity.
The $95\%$ CL limit for signal is defined as
$
\CLS=\CLSB/\CLB,
$
requiring $\CLS=0.05$.

The following sources of systematic uncertainty
affecting the normalization of the signal are taken into account:
The systematic uncertainty on the luminosity 
is estimated to be $6.5\%$.
The total uncertainty on the efficiency
amounts to $5\%$, and is 
dominated by the uncertainties on the
efficiency to reconstruct an isolated muon and on 
the trigger efficiency.
The uncertainty on the NLO $\Hpm$ production cross section from
choice of parton distribution functions and 
renormalization and factorization scales
is about $10\%$~\cite{bib-spira}. 
The uncertainty on the background from MC
is $27\%$ (Table~\ref{tab-like}). 
This includes the statistical uncertainty
and the systematic uncertainty 
on the measured $\bbbar$ cross section~\cite{bib-run1}.

The systematic uncertainties on signal and background are taken into account 
in the limit calculation through averaging over possible values of  
signal and background, as given by their 
probability distributions, which are assumed to be Gaussian~\cite{bib-tom}.
This procedure weakens the limit on the mass by about 1~GeV$/c^2$.
Other sources of systematic uncertainties,
such as interpolation procedure used for the
cross sections and the description of the mass resolution,
were examined and found to be negligible.

Figure~\ref{fig-cl} shows $\CLS$ as a function
of the mass of a doubly-charged Higgs boson.
The median expected $\CLS$
indicates the sensitivity of the experiment for our luminosity, 
assuming that there is no signal.
Taking into account systematic uncertainties, 
a lower mass limit of $118.4$~GeV$/c^2$ is obtained for a left-handed 
and $98.2$~GeV$/c^2$ for a right-handed doubly-charged Higgs boson,
assuming $100\%$ branching into muon pairs, hypercharge $Y=|2|$, and
Yukawa couplings $h_{\mu\mu}>10^{-7}$.
This is the first search for doubly-charged
Higgs bosons at hadron colliders. 
It significantly extends the previous mass limit~\cite{bib-OPAL} 
for a left-handed doubly-charged Higgs boson.

%
We thank the staffs at Fermilab and collaborating institutions, 
and acknowledge support from the 
Department of Energy and National Science Foundation (USA),  
Commissariat  \` a L'Energie Atomique and 
CNRS/Institut National de Physique Nucl\'eaire et 
de Physique des Particules (France), 
Ministry of Education and Science, Agency for Atomic 
   Energy and RF President Grants Program (Russia),
CAPES, CNPq, FAPERJ, FAPESP and FUNDUNESP (Brazil),
Departments of Atomic Energy and Science and Technology (India),
Colciencias (Colombia),
CONACyT (Mexico),
Ministry of Education and KOSEF (Korea),
CONICET and UBACyT (Argentina),
The Foundation for Fundamental Research on Matter (The Netherlands),
PPARC (United Kingdom),
Ministry of Education (Czech Republic),
Natural Sciences and Engineering Research Council and 
WestGrid Project (Canada),
BMBF (Germany),
A.P.~Sloan Foundation,
Civilian Research and Development Foundation,
Research Corporation,
Texas Advanced Research Program,
and the Alexander von Humboldt Foundation.
%

\end{document}